# Suppression of Antiferroelectric State in NaNbO$_3$ at High Pressure from In Situ Neutron Diffraction

**S. K. Mishra, M. K. Gupta, R. Mittal and S. L. Chaplot**

Solid State Physics Division, Bhabha Atomic Research Centre, Mumbai 400085, India.

**Thomas Hansen**

Institut Laue-Langevin, BP 156, 38042 Grenoble Cedex 9, France.

## Abstract

We report direct experimental evidence of antiferroelectric to paraelectric phase transition under pressure in NaNbO$_3$ using neutron diffraction at room temperature. The paraelectric phase is found to stabilize above 8 GPa and its crystal structure has been determined in orthorhombic symmetry with space group *Pbnm*. The variation of the structural parameters of the both orthorhombic phases as a function of pressure was determined. We have not found evidence for structural phase transition around 2 GPa as previously suggested in the literature based on Raman scattering experiments, however, significant change in Nb-O-Nb bond angles are found around this pressure. The response of the lattice parameters to pressure is strongly anisotropic with a largest contraction along <100>. The structural phase transition around ~ 8 GPa is followed by an anomalous increase in the orthorhombic strain and tilt angle associated with the R point ($q$= ½ ½ ½). Ab-initio calculation of the enthalpy in the various phases of NaNbO$_3$ is able to predict the phase transition pressure well.

**PACS**: 64.60.Ej, 61.05.F-, 77.80.-e, 77.84.Ek

**Introduction**

The current surge of interest in modeling, design, and synthesis of novel materials showing multi-functionality is fuelled by both potential technological applications and the underlying new physics [1-5]. Compounds with the perovskite structure represent an appealing class of novel materials that exhibits a variety of interesting physical and chemical properties, like: ferro- and antiferro-electricity, magnetism, multiferroic behaviour, superconductivity, ionic conductivity and a range of dielectric responses *etc* [1-11]. These materials are widely used for technological applications because of the physical properties that arise as a result of structural phase transitions that distort the ideal cubic perovskite structure. Under ambient conditions many ($ABO_3$) perovskites exhibit structures with low symmetries that can be derived from the cubic aristotype structure through the tilting and distortions of the $BO_6$ octahedra, and displacements of both the A and B cations from the centre of their coordination polyhedra. The type and magnitude of these distortions govern the nature of the physical properties observed. Therefore, a understanding of the microscopic origin which govern the structural state of a given perovskite, and how that state may change with external stimuli like temperature, pressure or composition, have long been the subject of investigation. Alkaline niobates (like potassium sodium niobate and lithium sodium niobate) with ultra-large piezoresponse comparable to $Pb(Ti_{1-x}Zr_x)O_3$ (PZT)) have evoked considerable interest as the next generation eco-friendly lead-free piezoceramics. $NaNbO_3$ is one of the most promising perovskite materials for both fundamental and applied research, as this material exhibits an unusual complex sequence of temperature driven structural phase transitions [8-9].

The hydrostatic pressure influence on ferroelectric (FE) and antiferrodistortive (AFD) phase transitions have received considerable attention for materials exhibiting perovskite structure [6,7]. In the perovskites, which are known displacive soft mode ferroelectrics, the stability of the ferroelectric phase (driven by zone center instability $q=0$) is reduced under pressure due to modification of the short-range interatomic and long range Coulomb interactions. On the other hand, field stability region of antiferrodistortive phase (driven by the zone boundary instabilities: $q≠0$) increases with increasing pressure. For example: pressure dependent Raman scattering experiment on pure $PbTiO_3$ showed that the soft mode (which is characteristic of ferroelectric tetragonal phase) vanishes about 12 GPa at 300 K. This correlates with ferroelectric tetragonal (P4mm) to cubic phase transition in $PbTiO_3$ [10]. In contrast of this, $SrTiO_3$, undergoes antiferrodistortive phase transition (from cubic to tetragonal (I 4/mcm)) at ~ 10 GPa and no other phase transition is observed on increase of pressure up to 53 GPa [11]. This clearly suggests that field stability region of antiferrodistortive phase increases with pressure. Understanding the interactions between these different instabilities is thus of fundamental importance. Thus, it is interesting to investigate the effect of pressure on sodium niobate which has competing zone centre and boundary instabilities simultaneously [8,9].

$NaNbO_3$ crystallizes in orthorhombic (antiferroelectric, space group: *Pbcm*) phase at ambient conditions and transform in rhombohedral (ferroelectric, space group: *R3c*) phase at 15 K. Based on our recent detailed temperature dependent neutron diffraction studies, the phase diagram of sodium niobate is presented [8,9]. The ferroelectric ordering can be stabilized by applications of external field, chemical substitution and reduction of particles size even at room temperature. Shen *et al* [12] found a soft mode in pure NaNbO$_3$ around ~ 42 cm$^{-1}$ with pressure followed by appearance and disappearance of Raman lines. They suggested that NaNbO$_3$ might undergo a first-order phase transition at 7 GPa and another significant structural change at 12 GPa, possibly a transition into the paraelectric phase. Shiratori *et al* [13-14] have shown that microcrystalline NaNbO$_3$, which stabilizes in an orthorhombic *Pbcm* structure at ambient pressure, showed bulk-like successive transitions at around 2, 6, and 9 GPa respectively. However, in our knowledge there are no reports of the structures of these crystallographic phases in the literature.

In this paper, we report our high pressure investigation of the structural phase transition in NaNbO$_3$ up to 11 GPa. We use the neutron diffraction technique, which is sensitive to oxygen atomic positions that are crucial in the phase transition in Perovskites. We provide evidence of antiferroelectric to paraelectric transition at about 8 GPa. We have successfully identified the transition pressure and the determine crystal structure upto 11 GPa. Theoretically, we have investigated phase stabilities region of various phases of NaNbO$_3$ with pressure using ab-intio DFT. Our calculation of the enthalpy of the various phases using ab-intio DFT is able to predict the phase transition and its pressure fairly well.

**Experiments**

Neutron powder diffraction experiments under pressure and room temperature were performed at the high-flux D20 diffractometer [15] with a Paris-Edinburgh (P-E) device [16] in the Institut Laue- Langevin, France. The "high resolution" mode (take-off angle 118º) was selected with a wavelength of 1.3594 Å (vertically focusing Germanium monochromator in reflection, (117) reflection). The sample, mixed with Pb metal as the pressure manometer, was loaded into an encapsulated Ti-Zr gasket filled with a 4:1 mixture of methanol-ethanol as a pressure medium before pressing in a P-E device. The structural refinements were performed using the Rietveld refinement program FULLPROF [17]. In all the refinements, the background was defined by a sixth order polynomial in 2θ. A Thompson-Cox-Hastings pseudo-Voigt with axial divergence asymmetry function was chosen to define the profile shape for the neutron diffraction peaks. Except for the occupancies of the atoms, all other parameters i.e., scale factor, zero correction, background and half-width parameters along with mixing parameters, lattice parameters, positional coordinates, and thermal parameters were refined. All the refinements have used the data over the full angular range of 3≤2θ≤ 126 degree; although in various figures only a limited range is shown for clarity.

**Result and Discussion**

Neutron diffraction patterns at room temperature were collected at fifteen pressures up to 11 GPa in pressure increasing cycle. Figure 1 illustrates the quality of the neutron diffraction patterns of the NaNbO$_3$ at selected pressure. NaNbO$_3$ has an antiferroelectric phase (*Pbcm*) at ambient condition. In general, an antiferroelectric phase consists of two or more sub-lattice polarizations of anti-parallel nature, which in turn gives rise to certain reflections in the diffraction pattern that, are labeled as superlattice reflections of the parent ideal cubic perovskite structure. The superlattice reflections around 2θ =32 and 48 degrees are among the strongest antiferroelectric peaks and are marked with arrows and labeled as S1 and S2 in fig. 1. The pressure dependence of neutron data show significant changes with pressure especially in terms of shifting of various peaks. The most prominent changes above 7 GPa have been observed in peaks around 2θ=32° which are shifted in opposite direction, while the peaks around 2θ=48° disappear. Detail Rietveld refinement of the powder diffraction data shows that diffraction patterns up to 6 GPa could be indexed using the orthorhombic structure (space group *Pbcm*). The Rietveld refinements proceeded smoothly, revealing a monotonic decrease in lattice constant and cell volume with increasing pressure. Attempts to employ the same orthorhombic structural model in the refinements at ~ 8 GPa (Fig. 2 (b)) proved unsatisfactory, and a progressive worsening of the quality of the Rietveld fits with increasing pressure was found. Fig. 2 (b) shows the diffraction data at 7.8 GPa could not be indexed with the orthorhombic phase (space group *Pbcm*). The most apparent signature of the subtle structural transformation that occurs at above 7 GPa is the inability of orthorhombic structure (space group *Pbcm*) to account satisfactory for the superlattice peaks around 32 degree. Thus, disappearance of peak around 2θ=48° and movement of peaks around 2θ=32° in opposite directions in powder neutron diffraction provide evidence for structural phase transitions in sodium niobate with pressure. In order to index additional superlattice reflection, we explored various possibilities and found that orthorhombic structure with space group *Pbnm* and cell dimensions √2×√2×2 (with respect to elementary perovskite cell) indexed all the reflections. This structure is identified as paraelectric in nature by its symmetry. The detailed structural parameters and goodness of fit for both the orthorhombic phases at room temperature and pressure P= 4.9 GPa and P= 11.2 GPa, as obtained from neutron diffraction data, are given in Table I. We have also collected few neutron diffraction data during pressure decreasing cycle. Analysis of the diffraction data suggests that the sample transformed to parent antiferroelectric phase (*Pbcm*) below 7 GPa (see right side panel of figure 2).

Figure 3 shows the pressure dependence of the structural parameters of NaNbO$_3$ at room temperature. For easy comparison, orthorhombic lattice parameters (a, b, and c) are converted using the relation $a_p = a/\sqrt{2}$, $b_p=b/\sqrt{2}$ and $c_p=c/4$ for *Pbcm* space group and $a_p = a/\sqrt{2}$, $b_p=b/\sqrt{2}$ and $c_p= c/2$ for *Pbnm* space group, respectively. It is clear from Fig. 3, the lattice parameter monotonically decrease in the entire range of our measurements. The response of the lattice parameters to pressure is strongly anisotropic with significantly largest contraction along <100>. Compressibility along a, b and c is 0.0027 GPa$^{-1}$, 0.0022 GPa$^{-1}$ and 0.0018

GPa$^{-1}$ respectively. Careful inspection of the lattice parameter along <100> shows change of slope at about 2, 6 and 8 GPa, respectively. The structural phase transition around ~ 8 GPa is followed by anomalous change in the pressure evolution of the orthorhombic strain [2(a-b)/(a+b) ratio] with pressure (see Fig. 4(c)). We noted that the *Pbnm* space group is not a subgroup of the *Pbcm* space group. This suggests that the transition is of first order in nature as also reported in Ca doped SrTiO$_3$ system [18] which undergoes orthorhombic *Pbcm* to *Pbnm* phase transition with x=0.40 (at room temperature). However, the pressure dependence of volume does not indicate any discontinuity. Pressure modifies the nature of phase transition [6], for example, pressure induced phase transition in PbTiO$_3$ (tetragonal to cubic phase) also does not show volume discontinuity, although the same transition driven by increase in temperature shows large volume discontinuity. Fitting the pressure versus volume data with a third-order Birch-Murnagham equation of state gives a bulk modulus B ($\approx$ 157.5 $\pm$ 1.0) GPa (with B′ =4 fixed) for the orthorhombic phase at room temperature. This experimental bulk modulus in NaNbO$_3$ is fairly close to that in KNbO$_3$ (B= 165 GPa) at room temperature.

Application of pressure modifies the structural parameters, such as the Nb–O bond length, Nb– O–Nb bond angles, and the distortion of the NbO$_6$ octahedra. The tilting of octahedra may be described in terms of bond angles of B-O- B (axial oxygen). We have determined the pressure-induced changes of atomic positions from the Rietveld refinements. Figure 4 shows that the averaged Nb-O bond length shrinks almost continuously whereas a small jump of the Nb-O-Nb bond angle can be clearly seen at 8 GPa. It is evident from the figure that with increasing pressure, Nb-O2-Nb bond angle first increases upto 2 GPa and then monotonically decreases upto 8 GPa. On the other hand Nb-O1-Nb bond angle sharply decreases upto 2 GPa and then it decreases monotonically with increasing pressure. Above 8 GPa, the bond angle increases with increasing pressure which may be associated with decreasing $\pi$ bond contribution [19].

In perovskite oxides the antiferrodistortive and ferroelectric instabilities have an opposite dependence on volume. Thus, the competition is quite sensitive to changes in the lattice constant by external stimuli like: pressure, stress or chemical substitution. Variation of tilt angle associated with R point (q= ½ ½ ½) which provide a measure of the distortion of octahedra, is calculated using the structural parameters and shown in figure 4(c). It clearly suggests that the tilt angle increases with increasing pressure with sharp enhancement above the phase transition pressure (~8 GPa). It could be noticed that the distortion in *Pbnm* phase is larger in comparison to that in *Pbcm* phase.

We recall the result of pressure dependent Raman scattering studies reported in literature [12, 13]. Shen *et al* [12] found a soft mode around ~ 42 cm$^{-1}$ with pressure followed by appearance and disappearance of certain Raman lines for pure NaNbO$_3$. They proposed that NaNbO$_3$ undergoes a first-order phase transition at 7 GPa and another significant structural change at 12 GPa, which was prospected as a transition into the paraelectric phase. It is consistent with our observations of phase transition at 8 GPa. However, Shiratori *et al* [12] have reported that microcrystalline NaNbO$_3$ showed bulk-like successive transitions at around 2, 6, and 9

GPa respectively. It is remarkable to notice that the Raman spectra did not show appearance or disappearance of Raman lines around 2 GPa. The conclusion was drawn on the basis of change in positions of Raman lines (slope; see fig. 2 of ref. 13) followed by enhancement of intensity for certain Raman lines. On the other hand, the transition at 6 GPa was supported by disappearance of Raman lines. In the present study, neutron diffraction patterns do not show appreciable change at 2 GPa, while prominent changes occur above 7 GPa. However, we find gradual changes in Nb-O- Nb bond angles around 2 GPa without any overall structural change. Thus, based on present study, we could conclude that sodium niobate undergoes structural transition [from *Pbcm* to *Pbnm*] about 8 GPa which is consistent with Raman scattering measurements. The observed anomaly in pressure dependent Raman spectra by Shiratori *et al* [13] around 2 GPa could be associated with change in Nb-O-Nb bond angles. Further, the orthorhombic structure (*Pbnm*) of sodium niobate is found stable upto 11 GPa in our studies.

The nature of bonding can be experimentally visualized via Fourier and charge density maps. Figure 5 (a-c and g-h) shows the nuclear scattering density map in ab –plane as calculated by Fourier transformation of neutron diffraction data at P= 0 & 10.3 GPa respectively at 300 K, using Fullprof [17]. In the antiferroelectric phase (space group: *Pbcm*), the asymmetric unit of the structure consists of two Na atoms, namely, Na1 at the 4c site (1/4+$u$,3/4,0) and Na2 at the 4d site (1/4+$u$,3/4 +$v$,1/4); one Nb atom at the 8e site (1/4+$u$,1/4+$v$,1/8 +$w$); and four O atoms, namely, O1 at the 4c site (1/4+$u$,1/4,0), O2 at the 4d site (1/4+$u$,1/4+$v$,1/4), O3 at the 8e site (1/2+$u$,0+$v$,1/8+$w$), and O4 at the 8e site (0+$u$,1/2 +$v$,1/8+$w$). On the other hand, the structure of the paraelectric phase (space group:*Pbnm*) has one Na atom at the 4c site (0+$u$,1/2 +$v$,1/4), Nb atom at the 4a site (0 0 0) and two O atoms, namely, O1 at the 4c site at (0+$u$,0 +$v$,1/4) and O2 at the 8e site (1/4+$u$,1/4 +$v$,0+$w$); in the asymmetric unit cell. In the ideal cubic $NaNbO_3$, Na and Nb atoms are surrounded by 12 and 6 neighbouring oxygen atoms, respectively, and all the bond distances between Na- O and Nb-O are equal. In the present case for both orthorhombic phases, as a result of displacement of Na, Nb and O atoms some of Na-O and Nb-O bonds become longer and remaining become shorter. It is evident from the figure that in *Pbcm* phase, in the Na1- O1 plane (Fig. 5(a)), Na1 shows anisotropic nature along <100>, while in the Na2-O2 plane (Fig. 5(b)), Na2 shows anisotropic along <010>. However, all the oxygen atoms show isotopic nature. Strong covalency has been observed in Nb-O3 &Nb-O4 bonds with short bond lengths of ~ 1.90 Å. In *Pbnm* phase, in the Na-O1 plane (Fig. 5(g)), both sodium and oxygen atoms show anisotropic nature. In the Nb-O2 plane (Fig. 5(h)), all the oxygen show isotropic nature.

We have also computed the charge density using the structural parameters obtained by Rietveld refinement of neutron diffraction data and VESTA software [20] and shown the same in Figure 5 (d-f and i-j) for both the orthorhombic phases. This clearly confirms that the two Nb-O bonds, namely Nb-O3 and Nb- O4 with short bond length have more covalent nature in antiferroelectric phase. In the paraelectric phase, all the Nb– O2 bonds (Fig. 5 (j) are isotropic in nature. Figure 5 (f and j) clearly reveals that charge is localized along the covalent Nb-O bonds, however, the Na-O bond appears to be an ionic bond in both the orthorhombic phases.

Although great effort has been devoted to understand the fundamental mechanism of the phase transitions observed in NaNbO$_3$ using first principle calculations [21-25], we still have little knowledge due to presence of several competing structural instabilities with very similar free energies. Theoretically, we can predict the transition pressure by plotting the free energy as a function of pressure for various phases and identifying the phases of minimum free energy. Theoretical calculations of the crystal structure, total energies, and dynamics require information about the interatomic forces which can be obtained either by using a quantum-mechanical *ab initio* formulation or using semi-empirical interatomic potentials. We have used the Vienna Ab-inito Simulation Package VASP-5.2 within generalized gradient approximation (PBE-GGA) [26-28]. The total energy calculations have been done using energy cutoff of 1100 eV and 8×8×8, 8×8×4 and 8×8×8 K points respectively for *R3c*, *Pbcm* and *Pbnm* phases, which are found to be sufficient for accuracy of the order of meV.

Figure 6 depicts the computed pressure dependence of the difference in the enthalpy $\Delta H = H_i - H_{cubic/R3c}$ of various phases of sodium niobate. For the lower pressure (Pc=3.3 GPa), only ferroelectric (*R3c*) phase has the lowest enthalpy ($\Delta H$), as it is the well-known ground state of NaNbO$_3$ at T=0 K. However at pressure above 3.3 GPa, the antiferroelectric phase (*Pbcm*) becomes favorable over the ferroelectric (*R3c*) phase. Further increasing the pressure above 10 GPa, the paraelectric phase (*Pbnm*) becomes favourable over the other two phases. This finding is consistent with the observation that the *Pbnm* phase has large distortion compared to *Pbcm* (see figure 4(c)). Thus, the calculation predict (figure 6) that sodium niobate may undergo successive phase transitions from ferroelectric to antiferroelectric (at 3.3 GPa) to paraelectric (at 10.5 GPa) phases. The phase transition pressure is fairly close to experimental values (8 GPa) for antiferroelectric to paraelectric phase.

**Conclusions:**

In conclusion, we have reported pressure dependent neutron diffraction data in NaNbO$_3$ show that the antiferroelectric phase transforms to a paraelctric phase at 8 GPa. We identify that the crystal structure of the paraelectric phase is orthorhombic with space group *Pbnm*. We have determined the bulk modulus of NaNbO$_3$ to be 158 GPa, which is fairly close to that observed in KNbO$_3$. Theoretically, we have also computed the enthalpy in various phase of sodium niobate and found that the orthorhombic structure with space group *Pbcm* could transform to the *Pbnm* structure at high pressure. The calculated phase transition pressure is fairly close to the experimental value. Additional high pressure experiments on sodium niobate at higher pressure are encouraged to see the stability region of *Pbnm* phase.

**Table I:** Structural parameters of NaNbO$_3$ obtained by Rietveld refinement of neutron diffraction at room temperature and (a) pressure= 4.9 GPa using *Pbcm* and (b) pressure= 11.2 GPa using *Pbnm* space groups respectively.

| (A) Pressure = 4.9 GPa Space Group: *Pbcm* | | | | | (B) Pressure = 11.2 GPa Space Group: *Pbnm* | | | | |
|---|---|---|---|---|---|---|---|---|---|
| Atoms Positional Coordinates and thermal parameter | | | | | Atoms Positional Coordinates and thermal parameter | | | | |
| | X | Y | Z | B(Å)$^2$ | | X | Y | Z | B(Å)$^2$ |
| Na1 | 0.2291(3) | 0.7500 | 0.0000 | 1.710(1) | Na | 0.0261(1) | 0.50556(1) | 0.2500 | 1.885(3) |
| Na2 | 0.2281(5) | 0.8019(7) | 0.2500 | 2.176(6) | Nb | 0.0000 | 0.0000 | 0.0000 | 1.213(1) |
| Nb | 0.2588(6) | 0.2624(2) | 0.1268(6) | 2.005(5) | O1 | -0.0688(3) | -0.0134(3) | 0.2500 | 1.512(2) |
| O1 | 0.3274(4) | 0.2500 | 0.0000 | 0.925(3) | O2 | 0.2058(4) | 0.2862(2) | 0.0476(4) | 1.174(4) |
| O2 | 0.1942(4) | 0.0557(6) | 0.2500 | 2.861(4) | | | | | |
| O3 | 0.5490(5) | 0.0355(4) | 0.1431(5) | 2.683(3) | | | | | |
| O4 | 0.9567(5) | 0.4721(6) | 0.1093(7) | 2.271(1) | | | | | |
| Lattice Parameters (Å) A= 5.4399(1) (Å); B= 5.5185(2) (Å) C= 15.3623(3) (Å); Volume = 461.18(5) (Å)$^3$ R$_B$= 1.73; R$_p$=19.1; R$_{wp}$=14.7; R$_{exp}$=5.75 $\chi^2$= 6.51 | | | | | Lattice Parameters (Å) A= 5.3491(1) (Å); B= 5.4537(1) (Å) C= 7.6525(2) (Å); Volume = 223.242(5) (Å)$^3$ R$_B$= 1.98; R$_p$=13.8; R$_{wp}$=12.4; R$_{exp}$=4.98 $\chi^2$= 6.17 | | | | |

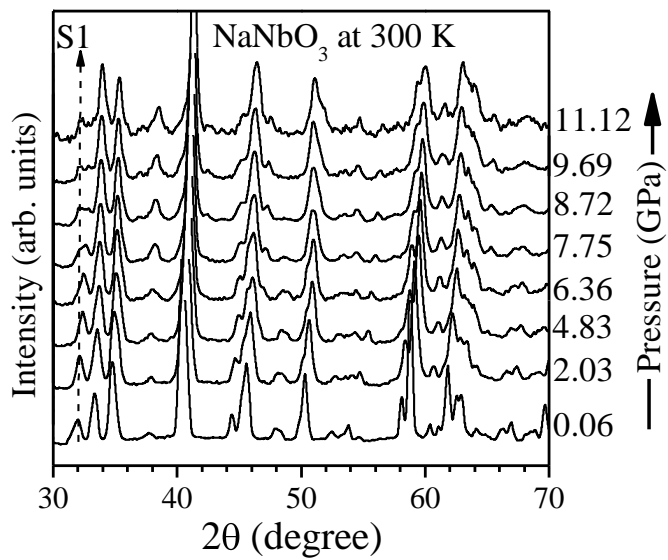

**Fig. 1** Evolution of the neutron diffraction patterns for NaNbO$_3$ at selected pressure. The characteristic superlattice reflections of antiferroelectric peak is marked with arrow and label as S1.

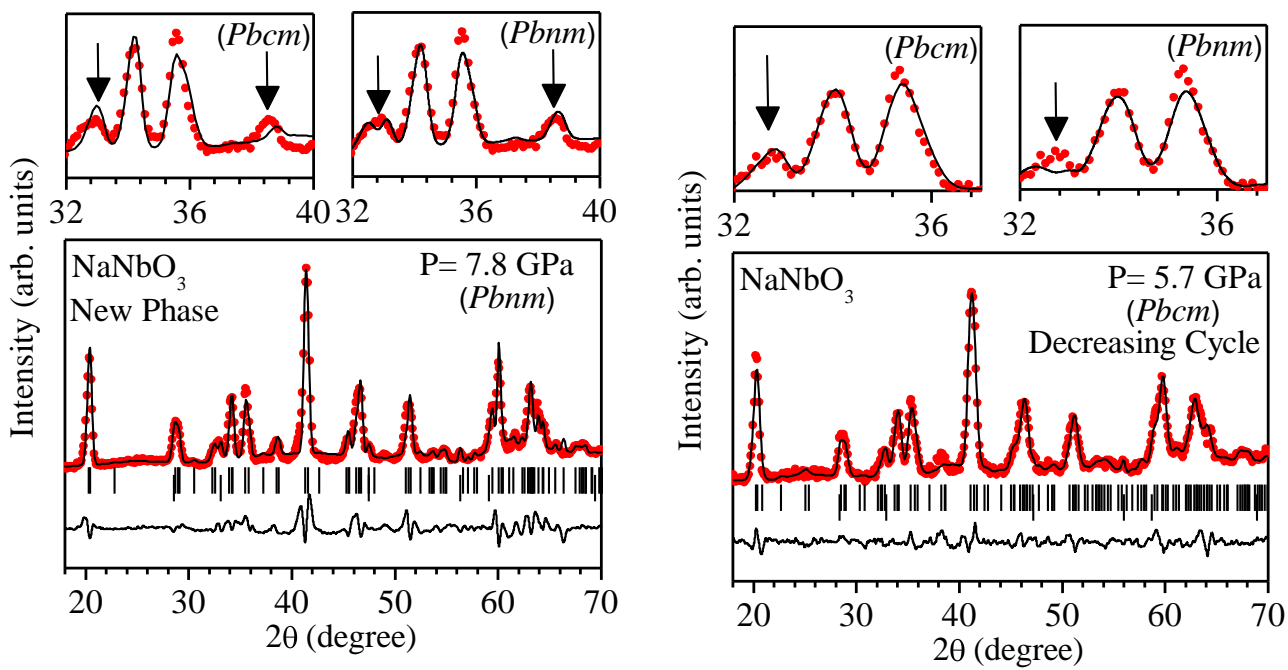

**Figure 2.** (color online) Observed (open circle), calculated (continuous line), and difference (bottom line) profiles obtained after the Rietveld refinement of NaNbO$_3$ using orthorhombic *Pbnm* space groups at 7.8 GPa pressure increasing cycle (left panel) and orthorhombic *Pbcm* space groups at 5.7 GPa during pressure decreasing cycle (right panel) at 300 K. The arrows show the non-accountabilities/accountabilities of the characteristic superlattice reflections using orthorhombic *Pbcm* and *Pbnm* space groups, respectively.

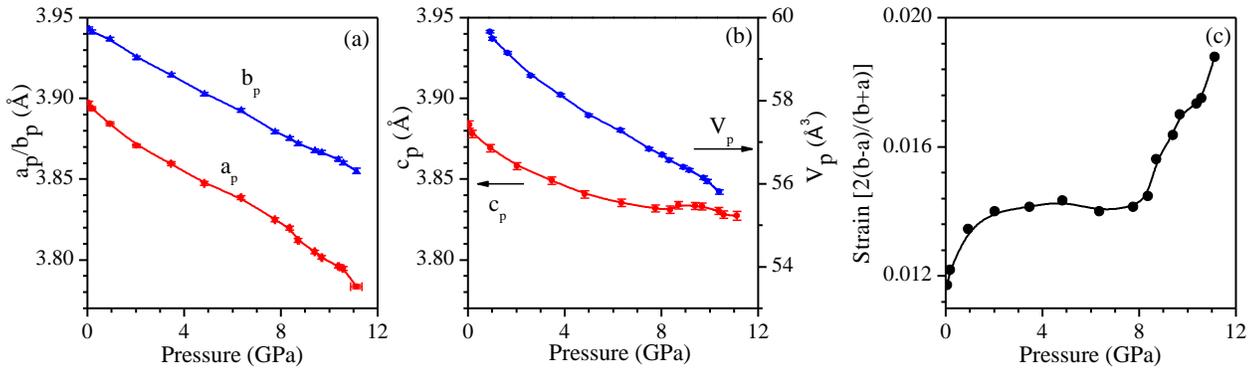

**Fig. 3** (color online) Evolution of the (a & b) structural parameters and (c) strain [2(b-a)/(b+a)] with pressure.

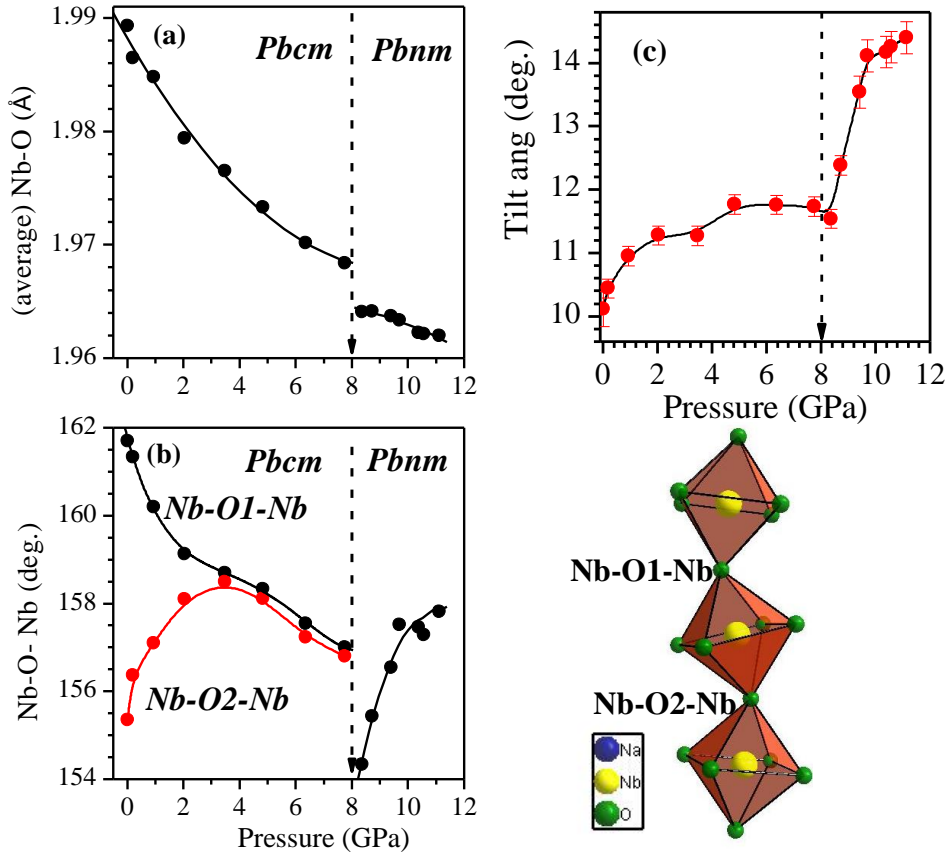

**Fig. 4** (color online) Variation of the (a) Nb-O bond length (average), (b) Nb-O-Nb bond angles and (c) tilt angle associated with R point (q= ½ ½ ½) for $NaNbO_3$ with pressure.

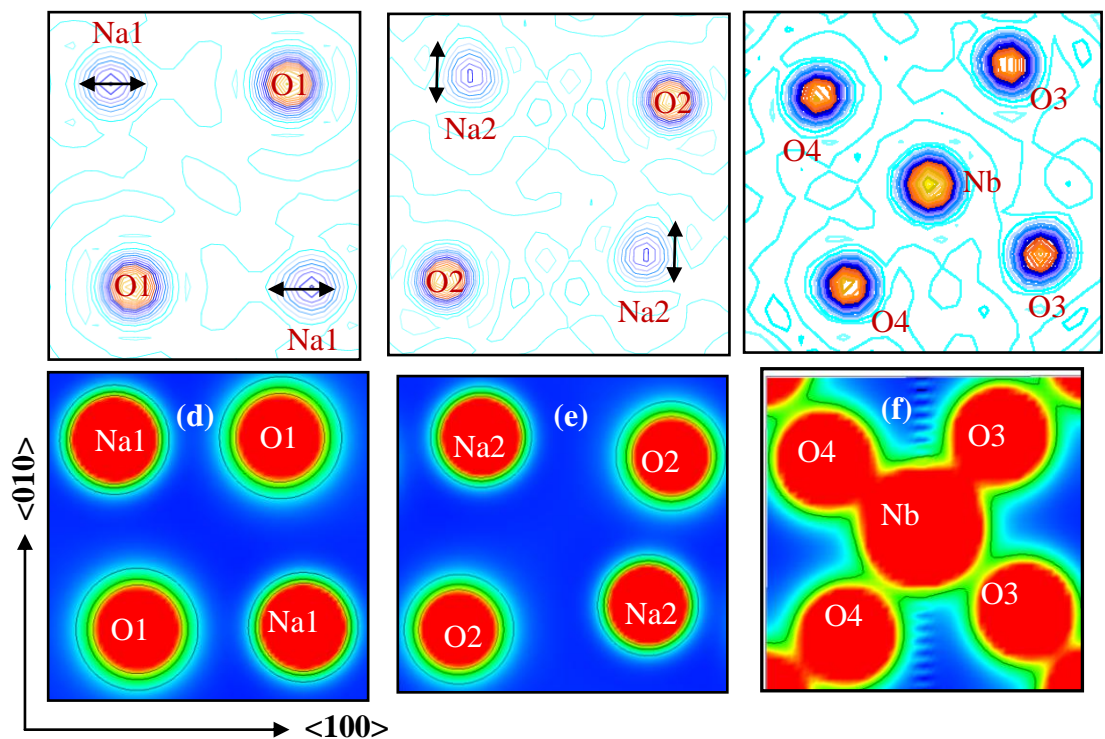
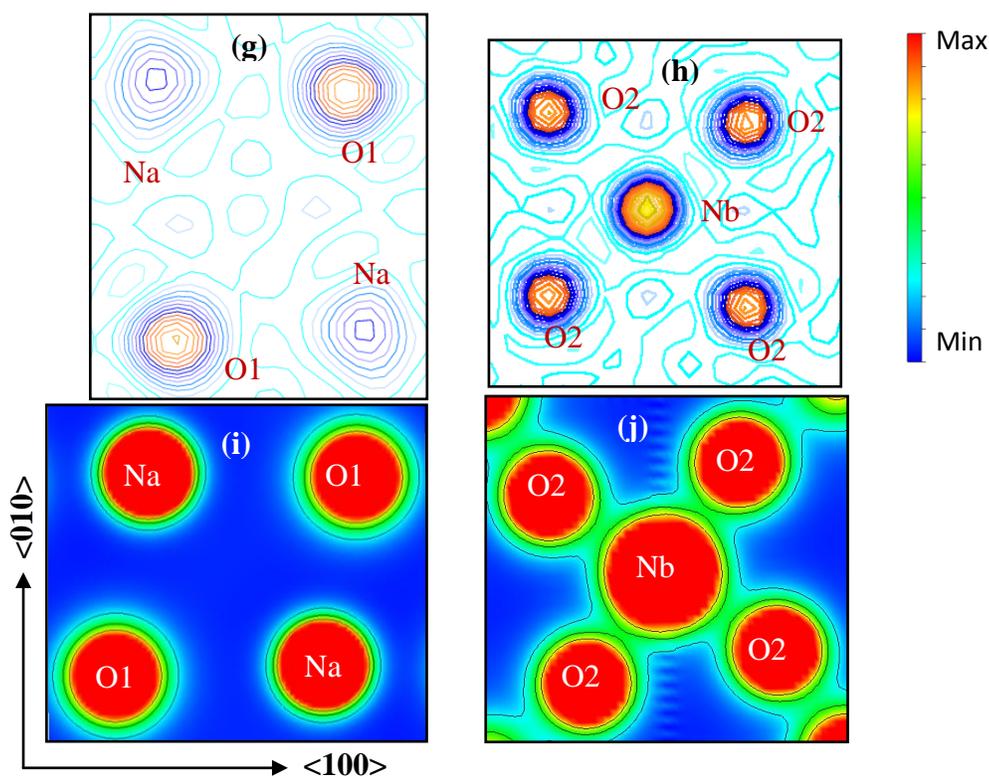

**Fig. 5** (color online) Computed nuclear scattering density map using Fourier transformation method of NaNbO$_3$ at P= 0 GPa (antiferroelectric phase with space group: *Pbcm*, panel: a-c) & ~ 10.4 GPa (paraelectric phase with space group: *Pbnm*, panel: g-h). The charge density computed by VESTA [20] for (d-f) for antiferroelectric and (i-j) for paraelectric phases respectively. The contour lines of the charge density are from 0 to 1.0 e Å$^{-3}$ with steps of 0.25 e Å$^{-3}$.

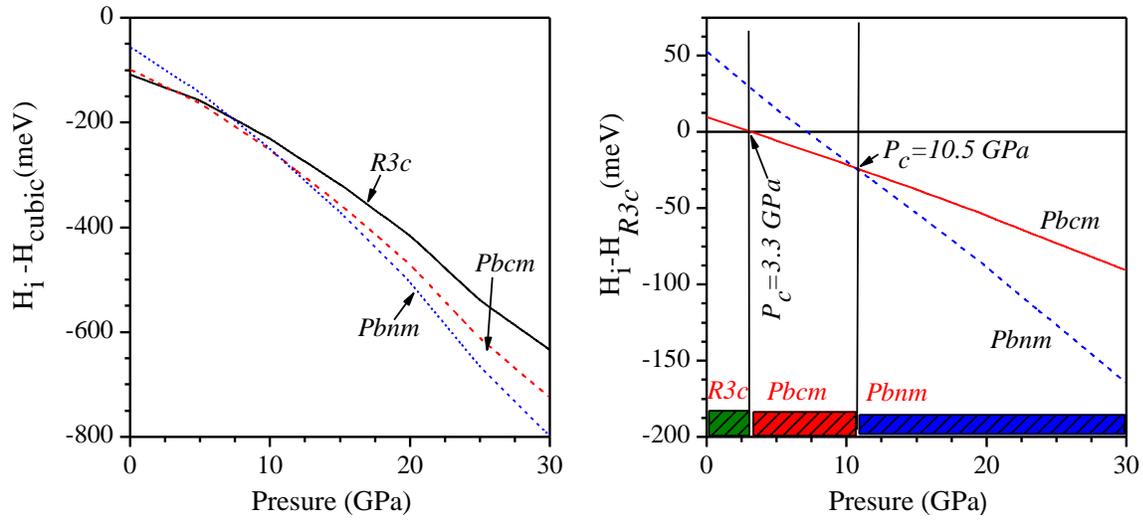

**Fig. 6.** (color online) Enthalpy difference (ΔH) between the indicated ferroelectric (*R3c*) antiferroelectric (*Pbcm*), paraelectric (*Pbnm*) and cubic (Pm3m) phases of $NaNbO_3$ as calculated using ab-inito DFT calculation. Right hand panel showed various phase stability region with pressure.